\magnification \magstep1
\raggedbottom
\openup 2\jot
\voffset6truemm
\def\cstok#1{\leavevmode\thinspace\hbox{\vrule\vtop{\vbox{\hrule\kern1pt
\hbox{\vphantom{\tt/}\thinspace{\tt#1}\thinspace}}
\kern1pt\hrule}\vrule}\thinspace}
\leftline {\bf ON THE PHOTON GREEN FUNCTIONS IN CURVED SPACE-TIME}
\vskip 1cm
\noindent
Giuseppe Bimonte, Enrico Calloni, Luciano Di Fiore, Giampiero Esposito,
Leopoldo Milano, Luigi Rosa
\vskip 1cm
\noindent
{\it Universit\`a di Napoli Federico II, Dipartimento
di Scienze Fisiche, Complesso Universitario di Monte S. Angelo,
Via Cintia, Edificio N', 80126 Napoli, Italy}
\vskip 0.3cm
\noindent
{\it Istituto Nazionale di Fisica Nucleare, Sezione
di Napoli, Complesso Universitario di Monte S. Angelo, Via 
Cintia, Edificio N', 80126 Napoli, Italy}
\vskip 1cm
\noindent
{\bf Abstract}.
Quantization of electrodynamics in curved space-time in the Lorenz
gauge and with arbitrary gauge parameter makes it necessary to
study Green functions of non-minimal operators with variable
coefficients. Starting from the integral representation
of photon Green functions, we link
them to the evaluation of integrals involving 
$\Gamma$-functions. Eventually, the full 
asymptotic expansion of the Feynman photon Green
function at small values of the world function, 
as well as its explicit dependence on the gauge 
parameter, are obtained without adding by hand a mass term to the
Faddeev-Popov Lagrangian. Coincidence limits of
second covariant derivatives of the associated
Hadamard function are also evaluated, as a first step towards the
energy-momentum tensor in the non-minimal case.
\vskip 100cm
\leftline {\bf 1. Introduction}
\vskip 0.3cm
\noindent
The investigation of Green functions has led, over many decades, to
several developments in quantum field theory [1--8]. In particular,
if flat Minkowski space-time is studied, we can say in modern language
that the bare photon Green function (or 
bare propagator) is obtained as follows:
\vskip 0.3cm
\noindent
(i) Work out the invertible gauge-field operator $P_{\mu \nu}$ 
acting on the potential $A^{\nu}(x)$, when the action functional 
consisting of Maxwell term plus gauge-fixing and ghost terms is
considered in the path integral.
\vskip 0.3cm
\noindent
(ii) Obtain the symbol $\Sigma(P_{\mu \nu})=\Sigma_{\mu \nu}(k)$
of $P_{\mu \nu}$ by working in momentum space. The symbol is a
$4 \times 4$ matrix of functions on the cotangent bundle of
${\bf R}^{4}$. To each $\partial_{\mu}$ term in the original operator
there corresponds ${\rm i}k_{\mu}$ in the symbol, while multiplicative
parts in $P_{\mu \nu}$ remain unaffected.
\vskip 0.3cm
\noindent
(iii) Invert the symbol $\Sigma_{\mu \nu}(k)$ to find the
matrix ${\widetilde \Sigma}^{\nu \lambda}(k)$ such that
$\Sigma_{\mu \nu}{\widetilde \Sigma}^{\nu \lambda}
={\widetilde \Sigma}^{\lambda \nu}\Sigma_{\nu \mu}
=\delta_{\mu}^{\; \lambda}$.
\vskip 0.3cm
\noindent
(iv) Obtain the photon Green function
$$
{\cal G}^{\mu \nu}(x,x')=\int_{\gamma}
{d^{4}k\over (2\pi)^{4}}{\widetilde \Sigma}^{\mu \nu}(k)
{\rm e}^{{\rm i}k \cdot (x-x')}
\eqno (1.1)
$$
where the choice of contour $\gamma$ reflects the choice of
boundary conditions.

In curved space-time, however, the gauge-field operator is
no longer a constant-coefficient partial differential operator,
and hence only a {\it local} momentum-space representation can 
be achieved, after using Riemann normal coordinates [7]. A
valuable alternative tool is instead provided by the space-time
covariant approach of DeWitt [2] to studying field theories possessing
infinite-dimensional invariance groups. With modern language, our
starting point is therefore an action functional $S$ consisting 
of the Maxwell term, plus gauge-fixing term in the Lorenz  
gauge\footnote {*}{this is due to L. Lorenz [9], not H. Lorentz}
plus ghost-field contribution, i.e. (hereafter $g \equiv -{\rm det} 
\; g_{\mu \nu}$)
$$
S=\int d^{4}x\sqrt{g}\left[-{1\over 4}g^{\mu \rho}g^{\nu \beta}
F_{\mu \nu}F_{\rho \beta}
-{(\nabla^{\mu}A_{\mu})^{2}\over 2\alpha}
-{\chi \over \sqrt{\alpha}} \cstok{\ }\psi \right]
\eqno (1.2)
$$
where $F_{\mu \nu} \equiv \nabla_{\mu}A_{\nu}
-\nabla_{\nu}A_{\mu}=\partial_{\mu}A_{\nu}
-\partial_{\nu}A_{\mu}$, $\nabla_{\mu}$ is the covariant derivative
with respect to the Levi-Civita connection, $\alpha$ is a
gauge parameter, $\chi$ and $\psi$ are independent ghost fields
obeying Fermi statistics [10]. More generally, the gauge-fixing
term [more appropriately called gauge averaging, from the
point of view of the path integral] for Maxwell theory might be
written in the form $-{\Phi^{2}(A)\over 2\alpha}$, with $\Phi$ any
functional on the space of (gauge) connection 1-forms
$A_{\mu}dx^{\mu}$ such that the resulting gauge-field operator
$P_{\mu \nu}$ on $A^{\nu}$ is invertible, but we here choose to
work in the Lorenz gauge, following Endo. The gauge parameter
$\alpha$ also occurs in the ghost action since, following Nielsen
and van Nieuwenhuizen [11], to the $\alpha$-dependent gauge-breaking
term there corresponds a non-trivial $\alpha$-dependence of the
ghost effective action.
The action (1.2) is
invariant under the BRST transformations for Abelian theory [8,12]
and can be eventually cast in the form
$$
S=\int d^{4}x \sqrt{g} \left[-{1\over 2}A_{\mu}P^{\mu \nu}(\alpha)
A_{\nu}+ {\chi \over \sqrt{\alpha}} P_{0} \psi \right]
\eqno (1.3)
$$
where the second-order differential operators $P^{\mu \nu}(\alpha)$
and $P_{0}$ read as
$$
P^{\mu \nu}(\alpha) \equiv -g^{\mu \nu}\cstok{\ }+R^{\mu \nu}
+\left(1-{1\over \alpha}\right)\nabla^{\mu}\nabla^{\nu}
\eqno (1.4)
$$
$$
P_{0} \equiv -\cstok{\ } \equiv -g^{\mu \nu}\nabla_{\mu}\nabla_{\nu}
\eqno (1.5)
$$
and we {\it assume} that they have no zero-modes.
Following the Fock--Schwinger--DeWitt method 
[2,13,14], we now consider two
abstract Hilbert spaces, spanned by basis vectors $|x \rangle$ and
$|x,\mu \rangle$ which satisfy the orthonormality conditions
$$
\langle x | x' \rangle =\delta(x,x')
\eqno (1.6)
$$
$$
\langle x,\mu | x',\nu \rangle = g_{\mu \nu}(x)\delta(x,x').
\eqno (1.7)
$$
Strictly, we should here refer to Gel'fand triples [15] rather than
abstract Hilbert spaces on their own, but we shall not be concerned
with this standard of mathematical rigour.
Of course, $|x \rangle$ is the familiar Dirac notation for the
eigenfunctionals of the position operator which has continuous
spectrum. They have a distributional nature and lead to a resolution
of the identity in the form
$\int d^{3}x |x \rangle \langle x |=I$. Moreover, the index of both
$|x, \mu \rangle$ and the associated `bra' $\langle x, \mu |$ is
viewed as that of a covariant vector density of weight ${1\over 2}$.
The `Hamiltonian' operators $H_{0}$ and $H(\alpha)$ associated to
$P_{0}$ and $P^{\mu \nu}$, respectively, are defined by (see
(1.5) and (1.4))
$$
\langle x | H_{0} | x' \rangle=P_{0} \langle x | x' \rangle
\eqno (1.8)
$$
$$
\langle x,\mu |H(\alpha)|x',\nu \rangle 
=P_{\mu}^{\; \lambda}(\alpha) 
\langle x,\lambda | x',\nu \rangle .
\eqno (1.9)
$$

The proper-time transformation kernels (also called `heat kernels'
by DeWitt [2,10], although in the mathematical literature one
speaks about heat kernels for elliptic operators on manifolds 
endowed with a positive-definite metric) are defined by
$$
K_{\mu \nu}^{(\alpha)}(x,x';\tau) \equiv 
\langle x,\mu | {\rm e}^{-{\rm i}\tau H(\alpha)}| x',\nu \rangle
\equiv K_{\mu \nu'}^{(\alpha)}(\tau)
\eqno (1.10)
$$
$$
K_{0}(x,x';\tau) \equiv \langle x | {\rm e}^{-{\rm i} \tau H_{0}}
| x' \rangle
\eqno (1.11)
$$
where $\tau$ is the proper-time parameter (not to be confused with the
Euclidean-time parameter of Euclidean field theory). The kernel
$K_{\mu \nu'}^{(\alpha)}(\tau)$ (where the index $\mu$ 
`lives' at $x$ and the index $\nu$ `lives' at $x'$) is a 
solution of the initial-value problem consisting of the
partial differential equation 
$$
{\rm i}{\partial \over \partial \tau}K_{\mu \nu'}^{(\alpha)}(\tau)
=P_{\mu}^{\; \lambda}(\alpha)K_{\lambda \nu'}^{(\alpha)}(\tau)
\eqno (1.12)
$$
subject to the initial condition
$$
K_{\mu \nu'}^{(\alpha)}(\tau=0)=g_{\mu \nu}(x)\delta(x,x').
\eqno (1.13)
$$
For arbitrary values of $\alpha$, the operator $P^{\mu \nu}$ 
in (1.4), as well as $P_{\mu}^{\; \lambda}=P^{\rho \lambda}g_{\rho \mu}$
in (1.12), is non-minimal, in that the wavelike-operator part
$-g^{\mu \nu}\cstok{\ }+R^{\mu \nu}$ is spoiled by
$\left(1-{1\over \alpha}\right)\nabla^{\mu}\nabla^{\nu}$.
Nevertheless, if one knows $K_{\mu \nu'}^{(\alpha)}$ at
$\alpha=1$, one can use this kernel, here denoted by
$K_{\mu \nu'}^{(1)}(\tau)$,
to evaluate $K_{\mu \nu'}^{(\alpha)}(\tau)$ according to 
the Endo formula [8]
$$
K_{\mu \nu'}^{(\alpha)}(\tau)=K_{\mu \nu'}^{(1)}(\tau)
+{\rm i} \int_{\tau}^{\tau / \alpha} dy
\nabla_{\mu} \nabla^{\lambda} K_{\lambda \nu'}^{(1)}(y).
\eqno (1.14)
$$

Equation (1.14) plays a key role in evaluating the regularized 
photon Green function (if no mass term or rotation of contour is
used, one needs a regularization for Green functions as well,
as we shall see following Endo [8]).
Sections 2 and 3 perform this general analysis, while the
full asymptotic expansion of the Feynman photon Green function
at small space-time separation of the points $x,x'$ is obtained
in section 4. Coincidence limits of 
second derivatives of the associated Hadamard function
are then evaluated in section 5, in light of their link with
the regularized energy-momentum tensor. Concluding remarks and
open problems are presented in section 6, while relevant details
are given in the appendices.
\vskip 0.3cm
\leftline {\bf 2. Photon Green functions in curved space-time}
\vskip 0.3cm
\noindent
As is well known, the photon Green function 
$G_{\mu \nu}^{(\alpha)}(x,x') \equiv G_{\mu \nu'}^{(\alpha)}$ 
(unlike the contravariant realization in
(1.1), we study hereafter the covariant one)
satisfies the differential equation
$$
\sqrt{g}P_{\mu}^{\; \lambda}(\alpha)G_{\lambda \nu}^{(\alpha)}(x,x')
=g_{\mu \nu}(x)\delta(x,x').
\eqno (2.1)
$$
The formal solution of Eq. (2.1) admits, in the absence of zero
and negatives modes, the integral representation
(here $g' \equiv -{\rm det} \; g_{\mu \nu}(x')$)
$$
g^{1\over 4}G_{\mu \nu'}^{(\alpha)}{g'}^{1\over 4}
={\rm i} \int_{0}^{\infty}d\tau \; K_{\mu \nu'}^{(\alpha)}(\tau)
\eqno (2.2)
$$
which is why we discussed the `heat-kernel' 
$K_{\mu \nu'}^{(\alpha)}(\tau)$ in the introduction. Bearing
in mind that integration in (2.2) is taken over the positive
half-line, we can say that it describes the {\it massless
limit of the Feynman propagator} (for which one would have
to add an infinitesimal negative imaginary mass). At this stage,
the formula (2.2) needs a suitable regularization, because we use
a Lorentzian-signature metric, so that contour rotation is not
exploited to obtain a convergent integral. Following
Endo, we use $\zeta$-function regularization and 
introduce a regularization parameter $\mu_{A}$ defining
(any suffix to denote regularization of the photon Green function
is omitted for simplicity of notation)
$$
g^{1\over 4}G_{\mu \nu'}^{(\alpha)}{g'}^{1\over 4}
\equiv \lim_{s \to 0}{\mu_{A}^{2s} \; 
{\rm i}^{s+1}\over \Gamma(s+1)}
\int_{0}^{\infty}d\tau \; \tau^{s} K_{\mu \nu'}^{(\alpha)}(\tau)
\eqno (2.3)
$$
where Eq. (1.14) should be used on the right-hand side of (2.3).
It should be stressed that the limit as $s \rightarrow 0$ should
be taken at the very end of all calculations, and cannot be brought
within the integral (2.3) (see section 4). 

The kernel $K_{\mu \nu'}^{(1)}(\tau)$ is known as 
$\tau \rightarrow 0$ and as $\sigma(x,x') \rightarrow 0$ through 
its Fock--Schwinger--DeWitt asymptotic expansion [2,13,14,16]
$$
K_{\mu \nu'}^{(1)}(\tau) \sim {{\rm i}\over 16 \pi^{2}}g^{1\over 4}
\sqrt{\bigtriangleup}{g'}^{1\over 4}
{\rm e}^{{\rm i} \sigma \over 2\tau}
\sum_{n=0}^{\infty}({\rm i}\tau)^{n-2}b_{n \; \mu \nu'}
\eqno (2.4)
$$
where $\sigma=\sigma(x,x')$ is half the square of the geodesic 
distance between $x$ and $x'$, the bi-scalar 
$\sqrt{\bigtriangleup}(x,x')$ is defined by the equation [10]
$$
\sqrt{g} \; \bigtriangleup \; \sqrt{{g'}}
={\rm det} \; \sigma_{; \mu \nu'}
\eqno (2.5)
$$
and the coefficient bi-vectors $b_{n \; \mu \nu'}$
are evaluated by solving a recursion formula obtained upon
insertion of (2.4) into Eq. (1.12). Such a recursion formula
reads, for all $n=0,1,2,...,\infty$, as 
$$
\sigma^{; \lambda}b_{n \; \mu \nu' ; \lambda}
+n b_{n \; \mu \nu'}
={1\over \sqrt{\bigtriangleup}}\left(\sqrt{\bigtriangleup} \;
b_{n-1, \mu \nu'}\right)_{; \lambda}^{\; \; \lambda}
-R_{\mu}^{\; \lambda} b_{n-1,\lambda \nu'}.
\eqno (2.6)
$$
For example, one therefore finds
$
b_{0 \; \mu \nu'}=g_{\mu \nu'},
$
which is the parallel displacement matrix [2,17] along the geodesic
between $x$ and $x'$.
The asymptotic expansion (2.4), called the local asymptotics 
of $K_{\mu \nu'}^{(1)}(\tau)$, only holds for 
$\tau \rightarrow 0$ at small values of $\sigma(x,x')$, which is
precisely the framework of interest in renormalization theory and
for applications to `laboratory physics'. 
At first sight, the above properties
suggest therefore splitting the integral 
$I_{\mu \nu'}(s,\alpha)$ in (2.3) into an integral
from $0$ to $b$ plus an integral from $b$ to $\infty$, where
$b \in ]0,1[$ to account for the $\tau \rightarrow 0$ limit.
However, as far as the asymptotic expansion of
$I_{\mu \nu'}(s,\alpha)$ is concerned, we can use (2.3), (1.14)
and integrate the local asymptotics (2.4) over the whole positive
half-line of $\tau$, 
when $x$ is very close to $x'$ (see appendix A). Hence we find
$$ 
I_{\mu \nu'}(s,\alpha) \equiv 
\int_{0}^{\infty}  
\tau^{s} K_{\mu \nu'}^{(\alpha)}(\tau) d\tau 
\sim I_{\mu \nu'}^{A}(s)
+I_{\mu \nu'}^{B}(s,\alpha)
\eqno (2.7)
$$
where
$$
I_{\mu \nu'}^{A}(s) \sim
{1\over 16 \pi^{2}}g^{1\over 4}\sqrt{\bigtriangleup}
{g'}^{1\over 4}\sum_{n=0}^{\infty}
b_{n \; \mu \nu'}{\rm i}^{n-1}F_{s,n}(x,x')
\eqno (2.8)
$$
$$
I_{\mu \nu'}^{B}(s,\alpha) \sim {1\over 16 \pi^{2}}
\sum_{n=0}^{\infty}{\rm i}^{n}\nabla_{\mu}\nabla^{\lambda}
\left(g^{1\over 4}\sqrt{\bigtriangleup}{g'}^{1\over 4}
b_{n \; \lambda \nu'}
{\widetilde F}_{s,n}(x,x';\alpha)\right).
\eqno (2.9)
$$
In the asymptotic expansions (2.8) and (2.9) we have defined
$$ 
F_{s,n}(x,x') \equiv 
(\sigma(x,x')/2)^{s+n-1}\int_{0}^{\infty}y^{-s-n}
{\rm e}^{{\rm i}y}dy
\eqno (2.10) 
$$
and
$$ 
{\widetilde F}_{s,n}(x,x';\alpha) \equiv
(\sigma(x,x')/2)^{n-1}\int_{0}^{\infty}d\tau \; \tau^{s} 
\int_{{\alpha \sigma(x,x') \over 2\tau}}^{{\sigma(x,x') \over 2\tau}}
y^{-n}{\rm e}^{{\rm i}y}dy. 
\eqno (2.11) 
$$
At this stage, the Feynman photon Green function has the
asymptotic expansion
$$
G_{\mu \nu'}^{(\alpha)}  \sim
g^{-{1\over 4}}\lim_{s \to 0}
{\mu_{A}^{2s}{\rm i}^{s+1}\over \Gamma(s+1)}
\Bigr(I_{\mu \nu'}^{A}(s)+I_{\mu \nu'}^{B}(s,\alpha)\Bigr)
({g'})^{-{1\over 4}}
\eqno (2.12)
$$
where the asymptotic expansions (2.8) and (2.9) should be used,
with integrals defined as in (2.10) and (2.11).
\vskip 0.3cm
\leftline {\bf 3. Regularized integrals} 
\vskip 0.3cm
\noindent
We are now going to evaluate the regularized integrals occurring
in the space-time covariant form 
of the Feynman photon Green function (for more
general results, see Refs. [18,19]). For this
purpose, we point out that the integral in the expression of
$F_{s,n}(x,x')$ (see (2.10)) is a particular case of the integral
$$
I(\beta) \equiv \int_{0}^{\infty}y^{-\beta}{\rm e}^{{\rm i}y}dy
={\rm i} \Gamma(1-\beta){\rm e}^{-{\rm i}{\pi \over 2}\beta}.
\eqno (3.1)
$$
Recall now that the $\Gamma$-function
$
\Gamma(z) \equiv \int_{0}^{\infty}y^{z-1}{\rm e}^{-y}dy,
$
originally defined on the half-plane ${\rm Re}(z)>0$, can be
analytically extended to a meromorphic function, 
with first-order poles 
at $0,-1,-2,...,-\infty$. 
With this understanding, we write that
$$
F_{s,n}(x,x')={\rm i}(\sigma(x,x')/2)^{s+n-1}\Gamma(1-s-n)
{\rm e}^{-{\rm i}{\pi \over 2}(s+n)}
\eqno (3.2)
$$
where $\Gamma(1-s-n)$ has first-order poles at $1-s-n=-k$, 
with $k=0,1,2,...,\infty$.

To evaluate the double integral occurring in (2.11), we first
exploit the identity 
$$ \eqalignno{
\int_{\alpha \sigma(x,x')\over 2\tau}^{\sigma(x,x')\over 2\tau}
y^{-n}{\rm e}^{{\rm i}y}dy & ={\rm i}^{3n+1}\left[\Gamma\left(1-n,
-{\rm i}{\alpha \sigma(x,x')\over 2\tau}\right) \right . \cr
& \left . -\Gamma 
\left(1-n,-{\rm i}{\sigma(x,x')\over 2\tau}\right)\right]
&(3.3)\cr}
$$
where in square brackets we have the incomplete $\Gamma$-function
$$
\Gamma(a,x) \equiv \int_{x}^{\infty}u^{a-1}{\rm e}^{-u}du.
\eqno (3.4)
$$
Hence we re-express ${\widetilde F}_{s,n}(x,x';\alpha)$ in the form
$$
{\widetilde F}_{s,n}(x,x';\alpha)=(\sigma(x,x')/2)^{n+1}{\rm i}^{3n+1}
\Bigr[I_{s,n}^{x,x'}(\alpha)-I_{s,n}^{x,x'}(1)\Bigr]
\eqno (3.5)
$$
where
$$
I_{s,n}^{x,x'}(\alpha) \equiv \int_{0}^{\infty}\tau^{s}
\Gamma \left(1-n,-{\rm i}{\alpha \sigma(x,x')\over 2\tau}\right)d\tau.
\eqno (3.6)
$$
At this stage, we are led to consider the integral
$$
J(\beta,\nu,c) \equiv \int_{0}^{\infty}x^{\beta-1}
\Gamma(\nu,cx)dx.
\eqno (3.7)
$$
On setting $y \equiv cx$, 
if ${\rm Re}(c)>0$, and exploiting the Leibniz rule and the
fundamental theorem of calculus one finds [20]
$$ 
J={1\over \beta c^{\beta}}\int_{0}^{\infty}
\left({d\over dy}y^{\beta}\right)
\left(\int_{y}^{\infty}u^{\nu-1}
{\rm e}^{-u}du \right)dy 
={\Gamma(\beta+\nu)\over \beta c^{\beta}}
\eqno (3.8) 
$$
because, for ${\rm Re}(\beta)>0$ and ${\rm Re}(\beta+\nu)>0$, 
the total derivative of $y^{\beta}\int_{y}^{\infty}
u^{\nu-1}{\rm e}^{-u}du$
yields vanishing contribution to (3.8).

We can however consider the analytic extension
of $\Gamma(\beta+\nu)$, after changing variable in the integral
(3.6) according to ${1\over \tau} \equiv T$, which yields
$$
I_{s,n}^{x,x'}(\alpha) \equiv \lim_{\varepsilon \to 0}
\int_{0}^{\infty}T^{-(s+2)}
\Gamma \left(1-n,\left(\varepsilon-{\rm i}{\alpha \sigma(x,x')\over 2}
\right)T \right)dT
\eqno (3.9)
$$
where a small positive $\varepsilon >0$ has been considered so as
to be able to apply the result (3.8). In our case, $\beta=-(s+1),
\nu=1-n,c=\varepsilon-{\rm i}{\alpha \sigma(x,x')\over 2}$, and after
making the analytic extension of $\Gamma(\beta+\nu)$ we find
$$
{\widetilde F}_{s,n}(x,x';\alpha)=-(\sigma(x,x')/2)^{s+n}
{\rm i}^{3(s+n)}{\Gamma(-s-n)\over (s+1)} (\alpha^{s+1}-1)
\eqno (3.10)
$$
where $\Gamma(-s-n)$ has first-order poles at $s+n=k$ for all 
$k=0,1,2,...,\infty$.
\vskip 0.3cm
\leftline {\bf 4. Asymptotic expansion of the Feynman photon Green function}
\vskip 0.3cm
\noindent
Our formulae (3.2) and (3.10) should be inserted into (2.8),
(2.9) and (2.12) to work out the full asymptotic expansion of
the Feynman photon Green function 
$G_{\mu \nu'}^{(\alpha)}$. For this
purpose, it is crucial to take the limit as $s \rightarrow 0$
in (2.12) at the last stage. 
Hence we find, as $x$ approaches $x'$ (which implies
$\sigma(x,x') \rightarrow 0$),
$$
G_{\mu \nu'}^{(\alpha)} \sim {{\rm i}\over 16 \pi^{2}}
\lim_{s \to 0}{\mu_{A}^{2s}\over \Gamma(s+1)}
{\cal G}_{\mu \nu'}^{(\alpha)}(s)
\eqno (4.1)
$$
where, after having defined
$$
U_{n \; \mu}^{\; \; \; \; \lambda}(s;\alpha) \equiv
{2 \over \sigma(x,x')}\delta_{\mu}^{\; \lambda}
+{(\alpha^{s+1}-1)\over (s+n)(s+1)}
\nabla_{\mu}\nabla^{\lambda}
\eqno (4.2)
$$
$$
B_{n \; \lambda \nu'}(s) \equiv 
b_{n \; \lambda \nu'}\sqrt{\bigtriangleup}(x,x')
(\sigma(x,x')/2)^{s+n}
\eqno (4.3)
$$
we write
$$
{\cal G}_{\mu \nu'}^{(\alpha)}(s) \equiv \sum_{n=0}^{\infty}
\Gamma(1-s-n)
U_{n \; \mu}^{\; \; \; \; \lambda}(s;\alpha) B_{n \; \lambda \nu'}(s).
\eqno (4.4)
$$

What is crucial for us is the $s \rightarrow 0$ limit
of the sum (4.4). Indeed, on studying first, for
simplicity, the case when the gauge-field operator reduces to a
minimal (wavelike) operator (i.e. at $\alpha=1$), one finds
$$
{\cal G}_{\mu \nu'}^{(1)}(s)={2 \sqrt{\bigtriangleup}(x,x')
\over \sigma(x,x')}\sum_{n=0}^{\infty}f_{n \; \mu \nu'}(s)
\eqno (4.5)
$$
having defined
$$
f_{n \; \mu \nu'}(s) \equiv 
\Gamma(1-s-n) b_{n \; \mu \nu'}
(\sigma(x,x')/2)^{s+n}.
\eqno (4.6)
$$
Since $b_{0 \; \mu \nu'}=g_{\mu \nu'}$ we therefore find
$$
{\cal G}_{\mu \nu'}^{(1)}(0)={2 \sqrt{\bigtriangleup}(x,x')
\over \sigma(x,x')}g_{\mu \nu'}
+{2 \sqrt{\bigtriangleup}(x,x')\over \sigma(x,x')}
\lim_{s \to 0}\sum_{n=1}^{\infty}f_{n \; \mu \nu'}(s)
\eqno (4.7)
$$ 
which is very encouraging, since the first term on the right-hand
side of (4.7) is precisely the first term in the Hadamard asymptotic
expansion at small $\sigma(x,x')$ [16]. On the other hand, the
Hadamard Green function is precisely the imaginary part of the
Feynman Green function, in agreement with our formula (4.1).
Eventually, we find therefore, at small $\sigma(x,x')$,
$$ \eqalignno{
\; & G_{\mu \nu'}^{(\alpha)} \sim {{\rm i} \over 8\pi^{2}}
{\sqrt{\bigtriangleup}(x,x')\over \sigma(x,x')+{\rm i} \varepsilon}
g_{\mu \nu'} \cr
&+{{\rm i}\over 16 \pi^{2}}\lim_{s \to 0}
\left[{(\alpha-1)\over s(s+1)}\nabla_{\mu}\nabla^{\lambda}
B_{0 \lambda \nu'}(s) \right . \cr
& \left . +\sum_{n=1}^{\infty} \Gamma(1-s-n)
U_{n \; \mu}^{\; \; \; \; \lambda}(s;\alpha)B_{n \; \lambda \nu'}(s)
\right]
&(4.8)\cr}
$$
i.e. the `flat' Feynman propagator, with $+{\rm i} \varepsilon$ term 
restored, plus corrections resulting from the gauge parameter
($\alpha \not =1$ leading to a non-minimal operator) and from
non-vanishing curvature.

A further crucial check is whether our infinite sum (4.4) is 
also able to recover the familiar $\log \sigma(x,x')$ singularity,
which occurs for massive theories in flat space-time, and, more
generally, even for {\it massless} theories (as is our case) but 
in {\it curved} space-time. For this purpose, it is enough to set
$\alpha=1$ and focus on the sum in (4.5), having defined 
$f_{n \; \mu \nu'}(s)$ as in (4.6). Such a sum can be studied 
with the help of the Euler--Maclaurin formula 
(see appendix A and Ref. [21]), which provides, among the others,
a term given by the integral (hereafter, since the discrete summation
index $n$ is replaced by the continuous variable $z$, we consider the
coefficients $b_{z \; \mu \nu'}$, functions of $z$ 
that reduce to the coefficient 
bi-vectors $b_{n \; \mu \nu'}$ for $z=n$)
$$ \eqalignno{
{J_{\mu \nu'}(s)\over \sqrt{\bigtriangleup}(x,x')} 
& \equiv \int_{0}^{\infty}
\Gamma(1-s-z)
b_{z \; \mu \nu'}(\sigma(x,x')/2)^{s+z-1}dz \cr
&=(\sigma(x,x')/2)^{s}\left \{
\int_{0}^{1}\Gamma(1-s-z)
b_{z \; \mu \nu'}
{\rm e}^{(z-1)\log(\sigma(x,x')/2)}dz \right . \cr
& \left . + \int_{1}^{\infty}
\Gamma(1-s-z) b_{z \; \mu \nu'}
{\rm e}^{(z-1)\log(\sigma(x,x')/2)}dz \right \}.
&(4.9)\cr}
$$
At this stage we set $z-1 \equiv y$ in the second integral
in curly brackets on the right-hand side of (4.9), which
therefore becomes
$$
{\widetilde J}_{\mu \nu'}(s)=\left(\int_{0}^{y^{*}}
+\int_{y^{*}}^{\infty}\right) \Gamma(-y-s)
b_{y+1, \mu \nu'}
{\rm e}^{y \log(\sigma(x,x')/2)}dy.
\eqno (4.10)
$$
We then begin to understand what happens: 
at small $\sigma(x,x')$, the integrand in (4.10) 
becomes exponentially damped, so that the resulting asymptotic
expansion of (4.10) is obtained from integration in the interval
$[0,y^{*}]$ for some $y^{*}$ in a small neighbourhood of the origin.
Here we first expand ${\rm e}^{y \log(\sigma(x,x')/2)}$ at small
$y$ {\it for fixed} $\sigma(x,x')$, 
and eventually take the $\sigma(x,x') \rightarrow 0$ limit.
Such a procedure yields the non-uniform asymptotic expansion
$$
{\widetilde J}_{\mu \nu'}(s) \sim 
\log(\sigma(x,x')/2) \int_{0}^{y^{*}}
y \Gamma(-y-s) b_{y+1, \mu \nu'}dy. 
\eqno (4.11)
$$
On taking the $s \rightarrow 0$ limit we therefore recover the
familiar $\log(\sigma(x,x'))$ singularity of the photon Green
function, which results from non-vanishing Riemann curvature
(in Minkowski space-time, the corresponding 
$b_{y+1, \mu \nu'}$ would instead vanish).
\vskip 0.3cm
\leftline {\bf 5. Second derivatives of the Hadamard function in the
coincidence limit}
\vskip 0.3cm
\noindent
In physical applications, one is interested in the 
energy-momentum tensor, which is obtained from the action
functional (1.2) as
$$
T^{\mu \nu} \equiv {2\over \sqrt{g}} 
{\delta S \over \delta g_{\mu \nu}}
=T_{\rm Maxwell}^{\mu \nu}+T_{\rm gauge}^{\mu \nu}
+T_{\rm ghost}^{\mu \nu}
\eqno (5.1)
$$
where (here $\alpha$ is the gauge parameter in (1.2),
cf. Ref. [16])
$$
T_{\rm Maxwell}^{\mu \nu} \equiv F_{\; \gamma}^{\mu}F^{\gamma \nu}
-{1\over 4}F^{\gamma \beta}F_{\gamma \beta}g^{\mu \nu}
\eqno (5.2)
$$
$$ 
\alpha T_{\rm gauge}^{\mu \nu} \equiv -A_{\; ; \beta}^{\beta \; \; \; \mu}
A^{\nu}-A_{\; ; \beta}^{\beta \; \; \; \nu}A^{\mu} 
+ \left[A_{\; ; \gamma \beta}^{\gamma}A^{\beta}
+{1\over 2}(A_{\; ; \gamma}^{\gamma})^{2}\right]g^{\mu \nu}
\eqno (5.3) 
$$
$$
T_{\rm ghost}^{\mu \nu} \equiv -\chi^{;\mu}\psi^{;\nu}
-\chi^{;\nu}\psi^{;\mu}
+\chi^{;\beta}\psi_{;\beta}g^{\mu \nu}.
\eqno (5.4)
$$
On considering 
the Hadamard Green function, which here equals the imaginary part
of the Feynman Green function, and is defined by
$$
G_{\mu \nu'}^{H} \equiv \langle \Bigr[A_{\mu},A_{\nu'}
\Bigr]_{+} \rangle 
\eqno (5.5)
$$
one therefore finds, for the regularized energy-momentum tensor, 
the decomposition [16]
$$
\langle T^{\mu \nu} \rangle 
= \langle T^{\mu \nu} \rangle_{\rm Maxwell}
+\langle T^{\mu \nu} \rangle_{\rm gauge}
+\langle T^{\mu \nu} \rangle_{\rm ghost}
\eqno (5.6)
$$
where
$$ \eqalignno{
\; & \langle T^{\mu \nu} \rangle_{\rm Maxwell}
={1\over 4} \lim_{x' \to x}\biggr[\Bigr(g^{\mu \rho}g^{\nu \tau}
-{1\over 4}g^{\rho \tau}g^{\mu \nu}\Bigr)g^{\gamma \beta} \cr
& \times \Bigr(G_{\gamma \beta';\rho \tau'}^{H}
+G_{\beta \gamma';\tau \rho'}^{H}
-G_{\gamma \tau';\rho \beta'}^{H} \cr
&-G_{\tau \gamma';\beta \rho'}^{H}
-G_{\rho \beta';\gamma \tau'}^{H}
-G_{\beta \rho';\tau \gamma'}^{H} 
+G_{\rho \tau';\gamma \beta'}^{H}
+G_{\tau \rho';\beta \gamma'}^{H}\Bigr)\biggr]
&(5.7)\cr}
$$
$$ \eqalignno{
\alpha \; & \langle T^{\mu \nu} \rangle_{\rm gauge}=\lim_{x' \to x}
\left[-{1\over 4}g^{\gamma \beta}E^{\mu \nu \; \rho \tau}
\Bigr(G_{\beta \tau';\gamma \rho'}^{H}
-G_{\tau \beta';\rho \gamma'}^{H}\Bigr) \right . \cr
& \left . +{1\over 8}g^{\gamma \beta}g^{\mu \nu}g^{\rho \tau}
\Bigr(G_{\beta \tau';\gamma \rho'}^{H}
+G_{\tau \beta';\rho \gamma'}^{H}\Bigr)\right]
&(5.8)\cr}
$$
$$
\langle T^{\mu \nu} \rangle_{\rm ghost}=\lim_{x' \to x}
\left[-{1\over 4}E^{\mu \nu \; \gamma \beta}
\Bigr(G_{;\gamma \beta'}^{H}+G_{;\beta \gamma'}^{H}\Bigr)\right],
\eqno (5.9)
$$
and we are assuming that our limits (5.7)--(5.9) do not depend
on the choice of vacuum state.
In the formula (5.8) we are using the DeWitt supermetric
$$
E^{\mu \nu \; \rho \tau} \equiv
g^{\mu \rho}g^{\nu \tau}+g^{\mu \tau}g^{\nu \rho}
-g^{\mu \nu}g^{\rho \tau}
\eqno (5.10)
$$
and in (5.9) we consider the ghost Hadamard function [16]
$$
G^{H}(x,x') \equiv \langle [\chi(x),\psi(x')]_{+}\rangle.
\eqno (5.11)
$$
We should now specify in which order the various operations we
rely upon are performed. Indeed, in the evaluation of the Feynman
Green function in section 4, we first sum over $n$ and then take the
$s \rightarrow 0$ limit. Here, we eventually obtain the
energy-momentum tensor of the quantum theory according to the
point-splitting procedure (5.6)--(5.9), {\it with the understanding
that the coincidence limit} $\lim_{x' \to x}$ {\it is the last
operation to be performed}. In general, the analytic continuation in
$s$ and the coincidence limit do not commute [22], but the 
point-splitting result for $T^{\mu \nu}$ can be made to agree with
the local $\zeta$-function method, as has been proved in detail
in Ref. [22] for scalar fields in curved space-time.

It is clear from (5.7) and (5.8) that our analysis of the 
energy-momentum tensor is virtually completed if we can provide a
closed expression for the coincidence limit
$
\lim_{x' \to x} G_{\gamma \beta'; \rho \tau'}^{H}.
$
For this purpose we point out that (4.1)--(4.4) and the coincidence
limits of appendix B show that the minimal-operator part of the
Hadamard function contributes the divergent part
(we write $\lim_{\varepsilon \to 0} \Gamma(\varepsilon-k)$, 
with $k=0,1,2,...$,
in the formulae for such divergent contributions,
where $\Gamma(\varepsilon-k)={1\over \varepsilon}
{(-1)^{k}\over k!}+{\rm O}(1)$)
$$
\lim_{\varepsilon \to 0}\left[\Gamma(\varepsilon) \left( 
[b_{1 \; \gamma \beta';\rho \tau'}]-{1\over 6}
[b_{1 \; \gamma \beta'}]R_{\rho \tau}\right)
-{1\over 2}\Gamma(\varepsilon-1)
[b_{2 \; \gamma \beta'}]g_{\rho \tau}\right]
$$
while the non-minimal operator part of the Hadamard function 
contributes further divergent terms given by (see (4.2)--(4.4))
$$
(\alpha-1)\lim_{\varepsilon \to 0}
\left[\Gamma(\varepsilon)(A_{\beta \gamma \rho \tau}
+B_{\beta \gamma \rho \tau})
+{1\over 2}\Gamma(\varepsilon-1)C_{\beta \gamma \rho \tau}\right]
$$
where (see appendix B) 
$$ \eqalignno{
A_{\beta \gamma \rho \tau} & \equiv \Bigr[
g_{\lambda \beta'; \; \; \gamma \rho \tau'}^{\; \; \; \; \; \; \lambda}
\Bigr]-{1\over 6}\Bigr[
g_{\lambda \beta'; \; \; \gamma}^{\; \; \; \; \; \; \lambda}\Bigr]
R_{\rho \tau} \cr
&+{1\over 6}\left(\Bigr[
g_{\lambda \beta'; \; \; \tau'}^{\; \; \; \; \; \; \lambda}\Bigr]
R_{\gamma \rho}-\Bigr[
g_{\lambda \beta'; \; \; \rho}^{\; \; \; \; \; \; \lambda}\Bigr]
R_{\gamma \tau} 
+ \Bigr[g_{\lambda \beta';\gamma \tau'}
\Bigr]R_{\; \rho}^{\lambda}
-\Bigr[g_{\lambda \beta';\gamma \rho}\Bigr]R_{\; \; \tau}^{\lambda}
\right) \cr
&+\left(\Bigr[g_{\lambda \beta';\rho \tau'}\Bigr]
{1\over 6}R_{\; \; \gamma}^{\lambda}
+\Bigr[\sqrt{\bigtriangleup}_{;\; \; \gamma \rho \tau'}^{\; \lambda}
\Bigr]g_{\lambda \beta}\right)
&(5.12)\cr}
$$
$$ \eqalignno{
B_{\beta \gamma \rho \tau}& \equiv {1\over 2}
\left(\Bigr[
b_{1 \; \lambda \beta'; \; \; \tau'}^{\; \; 
\; \; \; \; \; \; \; \lambda}
\Bigr]g_{\gamma \rho}-\Bigr[
b_{1\; \lambda \beta'; \; \; \rho}^{\; \; \; 
\; \; \; \; \; \; \lambda}
\Bigr]g_{\gamma \tau}\right) \cr
&-{1\over 2}g_{\rho \tau}\left(\Bigr[
b_{1\; \lambda \beta'; \; \; \gamma}^{\; \; \; 
\; \; \; \; \; \; \lambda}
\Bigr]+{1\over 6}\Bigr[b_{1 \; \lambda \beta'}\Bigr]
R_{\; \; \gamma}^{\lambda}\right) \cr
&-{1\over 12}\left(\Bigr[b_{1 \; \lambda \beta'}\Bigr]
(R_{\; \; \tau}^{\lambda}g_{\gamma \rho}
+R_{\; \; \rho}^{\lambda}g_{\gamma \tau})
+\Bigr[b_{1 \; \tau \beta'}\Bigr]R_{\gamma \rho}
+\Bigr[b_{1 \; \rho \beta'}\Bigr]R_{\gamma \tau} \right) \cr
&+{1\over 2}\left(\Bigr[
b_{1\; \rho \beta';\gamma \tau'}\Bigr]
-\Bigr[b_{1 \; \tau \beta';\gamma \rho}\Bigr]\right) \cr
&-{1\over 6}\left(-\Bigr[
b_{1\; \lambda \beta'}\Bigr]
(R_{\; \; \rho \gamma \tau}^{\lambda}
+R_{\; \; \tau \gamma \rho}^{\lambda})
+{1\over 2}\Bigr[b_{1 \; \gamma \beta'}\Bigr]
R_{\rho \tau}\right)
&(5.13)\cr}
$$
$$
C_{\beta \gamma \rho \tau} \equiv -2 \left(\Bigr[
b_{2 \; \tau \beta'}\Bigr]g_{\gamma \rho}
+\Bigr[b_{2 \; \rho \beta'}\Bigr]g_{\gamma \tau}
\right)-{1\over 2}\Bigr[b_{2 \; \gamma \beta'}
\Bigr]g_{\rho \tau}.
\eqno (5.14)
$$
We have therefore provided a covariant isolation of divergences
resulting from every coincidence limit of second covariant
derivatives of the Hadamard Green function. In manifolds without
boundary, the work in Ref. [8] suggested that the trace anomaly
resulting from the regularized $T^{\mu \nu}$ has the coefficient
of the $\cstok{\ }R$ term which depends on the gauge parameter
$\alpha$ (see, however, comments in section 6, end of
second paragraph therein).
In manifolds with boundary, the integration of such a
total divergence does not vanish, and further boundary invariants 
contribute to the regularized energy-momentum tensor.
Thus, the calculation expressed by (5.12)--(5.14) is not of
mere academic interest, but is going to prove especially useful
when boundary effects are included (cf. results in Ref. [23]).
Of course, physical predictions are expected to be independent
of $\alpha$, but the actual proof is then going to be hard.

More precisely, the work by Brown and Ottewill [24], which differs
from our approach because the $\alpha=1$ case is there considered
and the ${1\over \sigma}$ and $\log(\sigma)$ singularities in the
propagator are there {\it assumed rather than derived}, has been
exploited by Allen and Ottewill [25] to show that, on using the 
Ward identity and the ghost wave equation, the energy-momentum
tensor is $\alpha$-independent {\it up to geometric terms}, i.e. up
to polynomial expressions of dimension length$^{-4}$ formed from
the metric, the Riemann tensor and its covariant derivatives. 
As far as we can see, our formulae (5.12)--(5.14) have precisely
such a nature, having dimension\footnote {*}{the coincidence limits
of $b_{1\; \mu \nu'}$ and $b_{2 \; \mu \nu'}$ have dimension
length$^{-2}$ and length$^{-4}$, respectively, each covariant
derivative has dimension length$^{-1}$, while Riemann, Ricci and
the scalar curvature each have dimension length$^{-2}$.}
length$^{-4}$ and being built from the metric
and Riemann with its covariant derivatives (see appendix B), 
and have the merit of providing an explicit
form of the general result of Allen and Ottewill [25].
Furthermore, the extension of these results to manifolds with
boundary is, to our knowledge, an open research problem, and is of
course relevant, for example, for the Casimir effect, which is
a boundary effect in the first place.
\vskip 0.3cm
\leftline {\bf 6. Concluding remarks}
\vskip 0.3cm
\noindent
By relying upon the regularized 
integrals (3.2) and (3.10), we have evaluated the asymptotic
expansion of the Feynman photon Green function
(see comments following (2.2)), here expressed
in the form (4.1)--(4.4). Such an expansion corresponds to the
singular part (i.e. divergent as $\sigma(x,x') \rightarrow 0$)
of the exact photon Green function [26,27].
We have endeavoured not to include mass terms because their 
addition `by hand' spoils the gauge invariance of the original
action nor is compatible with BRST invariance.
Moreover, their addition
to the Lagrangian of spinor electrodynamics leads to a photon
propagator with a $k^{0}$ part in momentum space,
incompatible with perturbative renormalizability; this is compensated
by adding an auxiliary vector field 
which spoils unitarity [28]. It is 
therefore rather important to study photon Green functions from
the point of view of massless QED theory. The calculational 
techniques used in the presence of a ${m^{2}\over 2}A_{\mu}A^{\mu}$
term in the Lagrangian [2,10,16] are then no longer available, which
is why our formulae (4.2)--(4.4) provide a novel way of expressing
the familiar singularities in the asymptotic expansion of the Feynman
photon Green function.

Our second original contribution is a covariant isolation of
divergences in the energy-momentum tensor resulting from every
coincidence limit of second covariant derivatives of the Hadamard
Green function, including the contribution of the gauge parameter
$\alpha$ which leads to a non-minimal operator on the potential
in the path integral. In Minkowski space-time it is well-known
how to deal with arbitrary $\alpha$ in the photon propagator [5],
and hence it is desirable to deal with arbitrary $\alpha$ also in
curved space-time. One then finds, 
as we have done in (5.12)--(5.14), an explicit
expression of the geometric terms of dimension length$^{-4}$ up to
which the regularized energy-momentum tensor is known to be 
$\alpha$-independent, according to the general analysis of Allen
and Ottewill [25]. The $\alpha$-dependence of the electromagnetic
trace anomaly found in Ref. [8] is instead an incorrect claim,
since the analysis of Nielsen and van Nieuwenhuizen [11] has shown
that the corresponding $\alpha$-dependence of the ghost effective
action cancels such an $\alpha$-dependence [11,25].

After the early work in Refs. [7,8,26,29], for example, there has
been recent work by other authors. More precisely, the work in Ref.
[18] has obtained the Euclidean Green function for an operator of
Laplace type (corresponding to the choice $\alpha=1$), while the
work in Ref. [19] has obtained results of very general nature,
including in particular a recursive algorithm which holds for
non-minimal operators ($\alpha \not = 1$ in our case) with 
positive-definite metrics. Heat-kernel asymptotics for non-minimal
operators has instead been worked out in detail by Gusynin and his
collaborators (see, for example, the work in Refs. [30,31]).

We should admit that we might have based all our
analysis of Green functions on Eq. (2.31) of Ref. [32], but our
approach in sections 2-4 is more directly related to the 
properties resulting from the Fock--Schwinger--DeWitt asymptotic
expansion in the case of minimal operators.
Casimir energies will eventually motivate the use of a modified
Schwinger--DeWitt ansatz for manifolds with boundary [33], along the 
lines of the work in Ref. [34]. The central object in Ref. [34] is
indeed the scalar Feynman Green function, which is a first step 
towards the photon Green function in curved space-time,
eventually including boundary surfaces, which are crucial in
the Casimir effect [35,36]. At that stage, our formulae 
(5.12)--(5.14) will be part of the general formula leading to
the regularized energy-momentum tensor.
Hopefully, such investigations 
will find application to the exciting new problem of Casimir
apparatuses in a (weak) gravitational field [34,37], where the 
regularized energy-momentum tensor of QED in curved space-time 
can be used to evaluate the force acting on the Casimir apparatus.
\vskip 0.3cm
\leftline {\bf Acknowledgments}
\vskip 0.3cm
\noindent
We are much indebted to Ivan Avramidi for scientific discussions
and correspondence, and for providing the argument 
in Eqs. (A1)-(A3) of appendix A. Comments by Joseph Buchbinder,
Klaus Kirsten and Diego Mazzitelli have also been helpful.
The work of G Bimonte and G Esposito has been partially supported
by the National Project {\it SINTESI 2002}. 
\vskip 0.3cm
\leftline {\bf Appendix A}
\vskip 0.3cm
\noindent
We are now aiming to discuss under which conditions one can use
(2.3), (1.14) and integrate the local asymptotics (2.4) over the
whole positive half-line of $\tau$, relying upon private
correspondence with Ivan Avramidi. For this purpose, we start 
with the better defined Euclidean theory, where the heat
kernel [38] has truly such a nature, and has the following 
asymptotic expansion in four dimensions:
$$
U(x,x';t) = (4\pi t)^{-2}{\rm e}^{-{\sigma(x,x')\over 2t}}
[P_{N}(x,x';t)+R_{N}(x,x';t)]
\eqno (A1)
$$
where $P_{N}$, of a polynomial nature, 
denotes the first $N$ terms and $R_{N}$ is the
exponentially small remainder, for which
$$
\lim_{t \to 0}t^{-N}R_{N}(x,x';t)=0.
\eqno (A2)
$$
Now by temporary addition of a finite mass term, the integral
$$
\int_{0}^{\infty}{\rm e}^{-tm^{2}}U(x,x';t)dt
\eqno (A3)
$$
becomes meaningful (strictly, for $m$ such that the second-order
differential operator ruling the field becomes positive-definite),
and the contributions of $P_{N}$ and $R_{N}$ can be integrated
separately. At this stage, the mathematical limit as 
$m \rightarrow 0$, and Wick rotation back to `real time', make it
possible to obtain the asymptotic expansion (2.10).

The Euler--Maclaurin formula [21] used in section 4 asserts that, if
$f: [0,\infty[ \rightarrow {\bf R}$ is a function having even-order
derivatives which are absolutely integrable on $(0,\infty)$,
then, for all $k=1,2,...,\infty$,
$$ \eqalignno{
\; & \sum_{i=0}^{k}f(i)-\int_{0}^{k}f(x)dx={1\over 2}[f(0)+f(k)] \cr
&+\sum_{s=1}^{m-1}{{\widetilde B}_{2s}\over (2s)!}
\Bigr[f^{2s-1}(k)-f^{2s-1}(0)\Bigr]+R_{m}(k)
&(A4)\cr}
$$
where ${\widetilde B}_{2s}$ are the Bernoulli numbers, and 
$R_{m}(k)$ is the remainder term, majorized by 
$$
|R_{m}(k)| \leq (2-2^{1-m})
{|{\widetilde B}_{2m}| \over (2m)!}
\int_{0}^{k}|f^{2m}(x)|dx.
\eqno (A5)
$$
\vskip 0.3cm
\leftline {\bf Appendix B}
\vskip 0.3cm
\noindent
In the course of deriving and further elaborating 
the formulae (5.12)--(5.14), we need
the following coincidence limits 
as $x' \rightarrow x$ [39,40],
here denoted by square brackets [...] as in Refs. [2,10,17]:
$$
[\sigma]=[\sigma_{;\mu}]=[\sigma_{;\mu'}]=0
\eqno (B1)
$$
$$
[\sigma_{;\mu \nu}]=[\sigma_{;\mu'\nu'}]=g_{\mu \nu}
=-[\sigma_{;\mu \nu'}]
\eqno (B2)
$$
$$
[\sigma_{;\lambda \mu \nu}]=[\sigma_{;\lambda \mu \nu'}]
=[\sigma_{;\lambda \mu' \nu'}]
=[\sigma_{;\lambda' \mu' \nu'}]=0
\eqno (B3)
$$
$$
[\sigma_{;\lambda \mu \nu \rho}]=-{1\over 3}
(R_{\lambda \nu \mu \rho}+R_{\lambda \rho \mu \nu})
=[\sigma_{;\lambda \mu \nu' \rho'}]
=-[\sigma_{;\lambda \mu \nu \rho'}]
\equiv S_{\lambda \mu \nu \rho}
\eqno (B4)
$$
$$
[\sigma_{;\mu \nu \lambda \tau \rho}]
={3\over 4}(S_{\mu \nu \lambda \tau;\rho}
+S_{\mu \nu \tau \rho;\lambda}
+S_{\mu \nu \rho \lambda;\tau})
\eqno (B5)
$$
$$
\Bigr[g_{\mu}^{\; \nu'}\Bigr]=\delta_{\mu}^{\; \nu}
\eqno (B6)
$$
$$
[g_{\mu \nu'; \alpha}]
=[g_{\mu \nu'; \alpha'}]=0
\eqno (B7)
$$
$$
[g_{\mu \nu';\alpha \beta}]
=-[g_{\mu \nu';\alpha \beta'}]
=-{1\over 2}R_{\mu \nu \alpha \beta}
\eqno (B8)
$$
$$
[g_{\mu \nu';\alpha \beta \gamma}]=-{1\over 3}
\Bigr(R_{\mu \nu \alpha \beta;\gamma}
+R_{\mu \nu \alpha \gamma;\beta}\Bigr)
\eqno (B9)
$$
$$ \eqalignno{
\; & [g_{\mu \nu';\alpha \beta \gamma \delta}]
+[g_{\mu \nu';\beta \alpha \gamma \delta}]
+[g_{\mu \nu';\gamma \alpha \beta \delta}]
+[g_{\mu \nu';\delta \alpha \beta \gamma}] \cr
&+{1\over 6}\biggr[\Bigr(R_{\; \beta \alpha \gamma}^{\lambda}
+R_{\; \gamma \alpha \beta}^{\lambda}\Bigr)
R_{\mu \nu \lambda \delta} 
+\Bigr(R_{\; \beta \alpha \delta}^{\lambda}
+R_{\; \delta \alpha \beta}^{\lambda}\Bigr)
R_{\mu \nu \lambda \gamma} \cr
&+\Bigr(R_{\; \gamma \alpha \delta}^{\lambda}
+R_{\; \delta \alpha \gamma}^{\lambda}\Bigr)
R_{\mu \nu \lambda \beta} 
+\Bigr(R_{\; \gamma \beta \delta}^{\lambda}
+R_{\; \delta \beta \gamma}^{\lambda}\Bigr)
R_{\mu \nu \lambda \alpha} \biggr]=0
&(B10)\cr}
$$
$$
[g_{\mu \nu';\alpha \beta \gamma \delta}]
-[g_{\mu \nu';\beta \alpha \gamma \delta}]
=-R_{\mu \nu \alpha \beta ; \gamma \delta}
-{1\over 2}R_{\; \nu \gamma \delta}^{\lambda}
R_{\mu \lambda \alpha \beta}
\eqno (B11)
$$
$$
[g_{\mu \nu';\alpha \beta \gamma \delta'}]
=-[g_{\mu \nu';\alpha \beta \gamma \delta}]
+[g_{\mu \nu';\alpha \beta \gamma}]_{;\delta}
\eqno (B12)
$$
$$
[\sqrt{\bigtriangleup}]=1 \; \; \; 
[\sqrt{\bigtriangleup}_{;\mu}]
=[\sqrt{\bigtriangleup}_{;\mu'}]=0
\eqno (B13)
$$
$$
[\sqrt{\bigtriangleup}_{; \mu \nu}]
={1\over 6}R_{\mu \nu}
=-[\sqrt{\bigtriangleup}_{;\mu \nu'}].
\eqno (B14)
$$
$$
\Bigr[\sqrt{\bigtriangleup}_{;\alpha \beta \gamma}\Bigr]
={1\over 12}(R_{\alpha \beta;\gamma}+R_{\alpha \gamma;\beta}
+R_{\beta \gamma;\alpha})
\eqno (B15)
$$
$$ \eqalignno{
\; & \Bigr[\sqrt{\bigtriangleup}_{;\alpha \beta \gamma \delta}
\Bigr]=-{1\over 8} \biggr \{ \Bigr[
\sigma_{\; \; \; \rho \alpha \beta \gamma \delta}^{\; \rho}
\Bigr]-{1\over 3}\Bigr(
R_{\alpha \rho}R_{\; \beta \gamma \delta}^{\rho}+
R_{\beta \rho}R_{\; \alpha \gamma \delta}^{\rho} \cr
&+R_{\gamma \rho}R_{\; \alpha \beta \delta}^{\rho}
+R_{\delta \rho}R_{\; \alpha \beta \gamma}^{\rho}\Bigr)
+{1\over 3}\Bigr(
R_{\alpha \rho}S_{\; \beta \gamma \delta}^{\rho} \cr
&+R_{\beta \rho}S_{\; \alpha \gamma \delta}^{\rho} 
+R_{\gamma \rho}S_{\; \alpha \beta \delta}^{\rho}
+R_{\delta \rho}S_{\; \alpha \beta \gamma}^{\rho}\Bigr)\cr
&-{2\over 9}(R_{\alpha \beta}R_{\gamma \delta}
+R_{\alpha \gamma}R_{\beta \delta}
+R_{\alpha \delta}R_{\beta \gamma}) \biggr \}
&(B16)\cr}
$$
$$
\Bigr[\sqrt{\bigtriangleup}_{;\alpha \beta \gamma \delta'}\Bigr]
=-\Bigr[\sqrt{\bigtriangleup}_{;\alpha \beta \gamma \delta}\Bigr]
+\Bigr[\sqrt{\bigtriangleup}_{;\alpha \beta \gamma}\Bigr]_{;\delta}
\eqno (B17)
$$
$$
\Bigr[b_{1 \; \mu \nu'}\Bigr]={1\over 6}Rg_{\mu \nu}-R_{\mu \nu}
\eqno (B18)
$$ 
$$ \eqalignno{
\; & \Bigr[b_{2 \; \mu \nu'}\Bigr]=-{1\over 6}R R_{\mu \nu}
+{1\over 6}\cstok{\ }R_{\mu \nu}
+{1\over 2}R_{\mu \rho}R_{\; \nu}^{\rho}
-{1\over 12}R_{\; \; \; \; \; \; \mu}^{\lambda \sigma \rho}
R_{\lambda \sigma \rho \nu} \cr
&+\left({1\over 72}R^{2}+{1\over 30}\cstok{\ }R
-{1\over 180}R^{\rho \sigma}R_{\rho \sigma}
+{1\over 180}R^{\rho \sigma \lambda \psi}
R_{\rho \sigma \lambda \psi}\right)g_{\mu \nu}
&(B19)\cr}
$$
$$
\Bigr[b_{1 \; \mu \nu';\rho}\Bigr]={1\over 4}
\Bigr(R_{;\rho}+2R_{\; \rho ; \lambda}^{\lambda}\Bigr)g_{\mu \nu}
-R_{\mu \nu;\rho}
-{1\over 3}R_{\mu \nu \; \; \rho;\psi}^{\; \; \; \; \psi}
\eqno (B20)
$$
$$ \eqalignno{
\; & \Bigr[b_{1 \mu \nu';\rho \omega}\Bigr]
=-{1\over 3}\Bigr[b_{1 \lambda \nu'}\Bigr]
R_{\mu \; \; \rho \omega}^{\; \; \lambda}
+{1\over 3} \biggr \{
-{1\over 36}R g_{\mu \nu}R_{\rho \omega}
-R_{\mu \nu ; \rho \omega} \cr
&+{1\over 2}R_{\mu}^{\; \; \lambda}R_{\lambda \nu \rho \omega}
-{1\over 12}R R_{\mu \nu \rho \omega}
-{1\over 6}R_{\; \; \rho}^{\lambda}R_{\mu \nu \lambda \omega} 
-{1\over 6}R_{\; \; \omega}^{\lambda}R_{\mu \nu \lambda \rho} \cr
&+g_{\mu \nu}g^{\lambda \psi}
\Bigr[\sqrt{\bigtriangleup}_{;\lambda \psi \rho \omega}\Bigr] 
+g^{\lambda \psi}\Bigr[g_{\mu \nu';\lambda \psi \rho \omega}\Bigr]
\biggr \}
&(B21)\cr}
$$
$$
\Bigr[b_{1 \mu \nu';\alpha \beta'}\Bigr]
=-\Bigr[b_{1 \mu \nu';\alpha \beta}\Bigr]
+\Bigr[b_{1 \mu \nu';\alpha}\Bigr]_{;\beta}.
\eqno (B22)
$$
As is stressed in Ref. [17], the possibility of taking such 
coincidence limits relies on the assumption that the space-time
points $x$ and $x'$ have a unique geodesic passing through them. 
This is the case if the points are close enough to one another, 
but there are physical instances where it does not hold. The
world function is then no longer single-valued, and the existence
of partial (or covariant) derivatives is not obvious {\it a priori}.
A {\it global} theory of the world function $\sigma$, covering such
singular cases, is very complicated and goes beyond our aims; we
therefore always assume that the geodesic passing through $x$ and
$x'$ is {\it unique} and that {\it partial derivatives exist} [17].
\vskip 0.3cm
\leftline {\bf References}
\vskip 0.3cm
\item {[1]}
Streater R F and Wightman A S 1964 {\it P.C.T., Spin and Statistics,
and All That} (New York: Benjamin)
\item {[2]}
DeWitt B S 1965 {\it Dynamical Theory of Groups and Fields}
(New York: Gordon and Breach); DeWitt B S 2003 {\it The Global
Approach to Quantum Field Theory} (Oxford: Oxford University Press)
\item {[3]}
Glimme J and Jaffe A 1987 {\it Quantum Physics, a Functional Integral
Point of View} (Berlin: Springer--Verlag)
\item {[4]}
Strocchi F 1993 {\it Selected Topics on the General Properties
of Quantum Field Theory} (Singapore: World Scientific)
\item {[5]}
Weinberg S 1996 {\it The Quantum Theory of Fields. I}
(Cambridge: Cambridge University Press)
\item {[6]}
Dowker J S and Critchley R 1976 {\it Phys. Rev.} D {\bf 13} 3224
\item {[7]}
Bunch T S and Parker L 1979 {\it Phys. Rev.} D {\bf 20} 2499
\item {[8]}
Endo R 1984 {\it Prog. Theor. Phys.} {\bf 71} 1366
\item {[9]}
Lorenz L 1867 {\it Phil. Mag.} {\bf 34} 287
\item {[10]}
DeWitt B S 1984 in {\it Relativity, Groups and Topology II}
eds B S DeWitt and R Stora (Amsterdam: North--Holland)
\item {[11]}
Nielsen N K and van Nieuwenhuizen P 1988 {\it Phys. Rev.} 
D {\bf 38} 3183
\item {[12]}
Becchi C, Rouet A and Stora R 1975 {\it Commun. Math. Phys.}
{\bf 42} 127
\item {[13]}
Fock V A 1937 {\it Iz. USSR Acad. Sci. (Phys.)}
{\bf 4--5} 551
\item {[14]}
Schwinger J 1951 {\it Phys. Rev.} {\bf 82} 664
\item {[15]}
Esposito G, Marmo G and Sudarshan E C G 2004
{\it From Classical to Quantum Mechanics}
(Cambridge: Cambridge University Press)
\item {[16]}
Christensen S M 1978 {\it Phys. Rev.} D {\bf 17} 946
\item {[17]}
Synge J L 1960 {\it Relativity: The General Theory}
(Amsterdam: North--Holland)
\item {[18]}
Avramidi I G 1998 {\it J. Math. Phys.} {\bf 39} 2889
\item {[19]}
Avramidi I G and Branson T P 2001 {\it Rev. Math. Phys.}
{\bf 13} 847
\item {[20]}
Prudnikov A P, Brychkov A Yu and Marichev O I 1986
{\it Integrals and Series, Vol. 2} (New York: Gordon and Breach)
\item {[21]}
Wong R 1989 {\it Asymptotic Approximations of Integrals}
(New York: Academic Press)
\item {[22]}
Moretti V 1999 {\it J. Math. Phys.} {\bf 40} 3843
\item {[23]}
Dalvit D A R and Mazzitelli F D 1997 {\it Phys. Rev.} D
{\bf 56} 7779
\item {[24]}
Brown M R and Ottewill A C 1986 {\it Phys. Rev.} D {\bf 34} 1776
\item {[25]}
Allen B and Ottewill A C 1992 {\it Phys. Rev.} D {\bf 46} 861
\item {[26]}
DeWitt B S and Brehme R W 1960 {\it Ann. Phys. (N.Y.)}
{\bf 9} 220
\item {[27]}
Fulling S 1989 {\it Aspects of Quantum Field Theory in Curved
Space} (Cambridge: Cambridge University Press)
\item {[28]}
Esposito G 2002 {\it Found. Phys.} {\bf 32} 1459
\item {[29]}
Adler S L, Liebermann J and Ng Y J 1977 {\it Ann. Phys. (N.Y.)}
{\bf 106} 279
\item {[30]}
Gusynin V P, Gorbar E V and Roman'kov V V 1991 {\it Nucl. Phys.}
B {\bf 362} 449
\item {[31]}
Gusynin V P and Gorbar E V 1991 {\it Phys. Lett.} B {\bf 270} 29
\item {[32]}
Avramidi I G 1995 {\it J. Math. Phys.} {\bf 36} 1557
\item {[33]}
McAvity D M and Osborn H 1991 {\it Class. Quantum Grav.} {\bf 8} 603
\item {[34]}
Caldwell R R 2002 `Gravitation of the Casimir Effect and the
Cosmological Non-Constant' (ASTRO-PH 0209312).
\item {[35]}
Esposito G, Kamenshchik A Yu and Kirsten K 1999 {\it Int. J. Mod. Phys.}
A {\bf 14} 281
\item {[36]}
Esposito G, Kamenshchik A Yu and Kirsten K 2000 {\it Phys. Rev.}
D {\bf 62} 085027
\item {[37]}
Calloni E, Di Fiore L, Esposito G, Milano L and Rosa L 2002
{\it Phys. Lett.} A {\bf 297} 328
\item {[38]}
Vassilevich D V 2003 {\it Phys. Rep.} {\bf 388} 279
\item {[39]}
Christensen S M 1976 {\it Phys. Rev.} D {\bf 14} 2490
\item {[40]}
Birrell N D and Davies P C W 1982 {\it Quantum Fields in
Curved Space} (Cambridge: Cambridge University Press)

\bye